\documentclass[reprint,twocolumn,secnumarabic,amssymb, nobibnotes, aps, prd, showpacs]{revtex4-1}

\usepackage{graphicx}
\usepackage{dcolumn}
\usepackage{bm}
\usepackage{slashed}
\begin{document}

\preprint{AIP/123-QED}

\title[]{Polarization in low energy kaon-hyperon interaction}
\author{M. G. L. Nogueira-Santos}
 \email{magwwo@gmail.com}
\author{C. C. Barros, Jr}%
 \email{barros.celso@ufsc.br}
\affiliation{ 
Departamento de F{\'{i}}sica, CFM, Universidade Federal de Santa Catarina\\ Florian{\'{o}}polis SC, CEP 88010-900, Brazil
}%


\begin{abstract}
In this paper, we study the low energy kaon-hyperon interaction considering effective chiral Lagrangians 
that include kaons, $\sigma$ mesons, hyperons and the corresponding 
resonances.
The scattering amplitudes are calculated and then we determine the angular 
distributions and polarizations. 
\end{abstract}

\maketitle
\section{Introduction}

The study of the polarization of particles in hadronic interactions has always been a 
challenging subject to be understood.
Since the discovery of the $\Lambda$ polarization in high energy inclusive processes by Bunce
\cite{bu}, that was a completely unexpected result as far as at the time the polarization 
effects were expected to decrease with the energy and disapear at high energies, many
experiments have been performed that confirmed these results and determined the polarizations
for other hyperons and anti-hyperons produced
in proton-nucleus \cite{hel}-\cite{mor} and 
also in heavy ion collisions \cite{RH1},\cite{plhc1}. Many models based in different physical theories
\cite{lund}-\cite{trosh} have been proposed in order to explain these results, always 
presenting some difficulties in this task. Recent data and models
keep this subject even more interesting \cite{star2}-\cite{mart}.
This fact shows that the polarization, due to its difficulty to be explained, is an observable 
that selects the theories that may be used, and then, is of great importance in order to 
improve the knowledge about the physical process that is being studied.   

Hyperon interactions is another subject of great importance in many physical systems. When
studying the hypernuclei structure for example \cite{hnuc1}-\cite{hnuc4}, probably the 
main problem is the determination of the nucleon-hyperon and hyperon-hyperon potentials
with enough accuracy, what has not been done yet. Fundamental results for this determination 
are the meson-nucleon and meson-hyperon interactions.

In the study of the hyperon stars structure \cite{hipstar1}-\cite{hipstar4}, the same
 problem occurs, and an accurate knowledge of the hyperon interactions is needed in order
to determine a suitable equation of state, and then the mass-radius relation for this
kind of star.

In order to study the hyperon and antihyperon polarizations that are observed in high  energy
proton-nucleus and in heavy ion 
collisions, a model based in the hydrodynamical aspects of the system has been proposed
\cite{cy}-\cite{ccb2}. In this model, after the collision, a hot expanding medium is produced,
and inside of it, as it is currently considered, the hyperons (and antihyperons) are produced.
These particles interact with the surrounding ones and then become polarized. Despite the 
fact that these particles are observed with high energies, the relative energy between
the particles inside the fluid is small, and then, the most important process to be considered
is the low energy meson-hyperon (and meson-antihyperon) interaction that determines the
cross-sections and polarizations for the particles that interact inside this fluid. This model
was able to describe the hyperon and antihyperon polarizations in many reactions 
\cite{cy}-\cite{ccb2} by considering the final state pion-hyperon interactions. 
A further improvement that must be done in the model is the inclusion of the 
kaon-hyperon and kaon-antihyperon interactions and verify this effect in the final results.
This kind of calculation is a way to investigate if the results of low energy collisions are
in accord
with the experimental data obtained in high energy  collisions.
For this reason a detailed study of the polarization that occurs in the low energy 
kaon-hyperon interactions is a fundamental aspect. 

So, the main objective of this paper is to calculate the low energy kaon-hyperon polarization.
This calculation will be based in a recently proposed model \cite{sant}, where the interaction
is described with effective lagrangians that include baryons, resonances and mesons as
degrees of freedom.

This paper will present the following content: in Sec. II the basic formalism will be revised,
in Sec. III the amplitudes of scattering will be calculated and in Sec. IV the results will be 
shown. In Sec. V the discussions and conclusions will be presented and some expressions
of interest will be shown in the Appendix.

\section{Polarization and Angular distribution formalism}

As it has been said before, in this paper we are interested in calculating the polarization 
that results from kaon-hyperon interactions at low energies. So, in this section  
we will present the basic formalism that may be used to calculate the observables of interest
 in terms  of partial wave decompositions of the scattering amplitudes
\cite{manc}, \cite{BH}, \cite{sant}.

The polarization and angular distribution are defined in terms of the $f$ and $g$ 
amplitudes as
\begin{equation}
\vec{P}=-2 \frac{Im(f^*g)}{|f|^2+|g|^2}\hat{n} 
\label{eq:}
\end{equation}
\noindent
and
\begin{equation}
\frac{d\sigma}{d\Omega}=|f|^2+|g|^2 \ ,
\label{eq:}
\end{equation}
\noindent
where $\hat n$ is a vector normal to the scattering plane, that is determined by the momenta
of the interacting particles. 
 $f(k,\theta)$ and $g(k,\theta)$ 
are the spin-non-flip and spin-flip amplitudes that may be written as
functions of the incident momentum $k$ and the scattering angle $\theta$ in the
center-of-mass frame. 

For the kaon-hyperon ($KY$) scattering $KY\rightarrow KY$
we define the amplitude $T_{K Y}$
\noindent
\begin{equation}
T_{K Y}=\sum_IT^IP_I \  ,
\label{eq5}
\end{equation}
\noindent
that is a sum over all the isospin $(I)$ states where
$P_I$ are the projector operators relative to these
states. The $T^I$  amplitudes may be parametrized as
\begin{equation}
T^I=\overline{u}(\vec{p'})\Big[A^{I}+\frac{1}{2}(\slashed{k}+\slashed{k}')B^I\Big]u(\vec{p})
\   ,
\label{eq:}
\end{equation}
where
 $u(\vec{p})$ is a spinor that represents the initial baryon incoming with four-momentum $p_\mu$, 
$p'_\mu$ is the outgoing baryon  four-momentum
and $k_\mu$ and $k_\mu '$ are the incoming and outgoing meson four-momenta.
In this work
the $A^I$ and $B^I$ amplitudes will be calculated from the Feynman diagrams.

For each isospin state the 
$f^I(k,\theta)$ and $g^I(k,\theta)$ amplitudes are given by the relation  
\begin{equation}
\frac{T^I}{8\pi \sqrt{s}} =
f^I(k,\theta)+ g^I(k,\theta)i \vec{\sigma}.\hat{n} \ ,
\end{equation}

\noindent
where $\sqrt{s}$ is the  total energy in the center-of-mass 
frame defined in the Appendix and
 may be  expanded in terms of unitarized partial-wave amplitudes $a_{l\pm}^{U}$
\begin{eqnarray}
 & &f^I(k,\theta)=\sum_{l=0}^\infty{\Big[(l+1)a_{l+}^{UI}(k)+la_{l-}^{UI}(k)\Big] P_l (\theta)}\ ,\\
 & &g^I(k,\theta)=i\sum_{l=1}^\infty{\Big[a_{l-}^{UI}(k)-a_{l+}^{UI}(k)\Big] P_l^{(1)} (\theta)}\ .
\end{eqnarray}

 The unitarization is necessary at the tree-level,
 these contributions are real and consequently 
violate the unitarity of the $S$ matrix. 
So, as it is usually done,we may reinterpret these results
as elements of the $K$ reaction matrix \cite{BH}, \cite{cm}, \cite{ccb}
and then obtain unitarized amplitudes
\begin{equation}
a_{l\pm}^U=\frac{a_\pm}{1-i|\vec{k}|a_\pm}   \ .
\end{equation}

The partial-wave amplitudes $a_{l\pm}^I(k)$
are calculated using the Legendre polynomials ($P_{l}(\theta)$) orthogonality relations
\begin{equation}
a_{l\pm}^I(k)=
\frac{1}{2}\int_{-1}^1\Big[P_l(\theta)f^I_1(k,\theta) +P_{l\pm 1}(\theta)f^I_2(k,\theta)\Big] d\theta \ ,
\end{equation}
with	
\begin{eqnarray}
f_1^{I}(k,\theta)&=&\frac{(E+m)}{8\pi \sqrt{s}}[A^{I}+(\sqrt{s}-m)B^I]\ ,\\
f_2^{I}(k,\theta)&=&\frac{(E-m)}{8\pi \sqrt{s}}[-A^{I}+(\sqrt{s}+m)B^I]\ ,
\end{eqnarray}
where $E$ is the baryon energy, $m$ is its mass (see the Appendix). 

At low energies the  $S$ ($l=$0) and $P$ ($l=$1) waves dominate the scattering amplitudes.
For higher values of $l$ they
are much smaller (almost negligible) and may be considered as small corrections.

In the next section we will calculate the amplitudes $A$ and $B$ using 
a model based in the chiral Lagrangian formalism for many $KY$ reactions. These amplitudes determined, the $T^I$ amplitudes will be determined and then we will be able to compute the polarization and the angular distribution.
 
\section{Scattering amplitudes}

In this section we will calculate the scattering amplitudes $A^I$ and $B^I$
by considering the effective nonlinear
chiral Lagrangians from \cite{manc}, \cite{BH}, \cite{sant} in the study
of the $K\Lambda$, $\overline K\Lambda$,  $K\Sigma$ and $\overline K\Sigma$
interactions.

Observing the fact that the $\Lambda$ hyperon
has isospin 0, there is just one isospin channel to be considered
in the $K\Lambda$ interaction
and then the scattering amplitude  will have the form
\begin{equation}
T_{K\Lambda}=\bar{u}(\vec{p'})\Big[A+\Big(\frac{\slashed k+\slashed k'}{2}\Big)B\Big]u(\vec{p})\ .
\label{eq20}
\end{equation} 
Comparing with (\ref{eq5}), we observe that in this interaction
we 
have a trivial formulation with $P_{1/2}=1$, as the kaon has isospin $1/2$.

\begin{figure}[!htb]
 \includegraphics[width=0.45\textwidth]{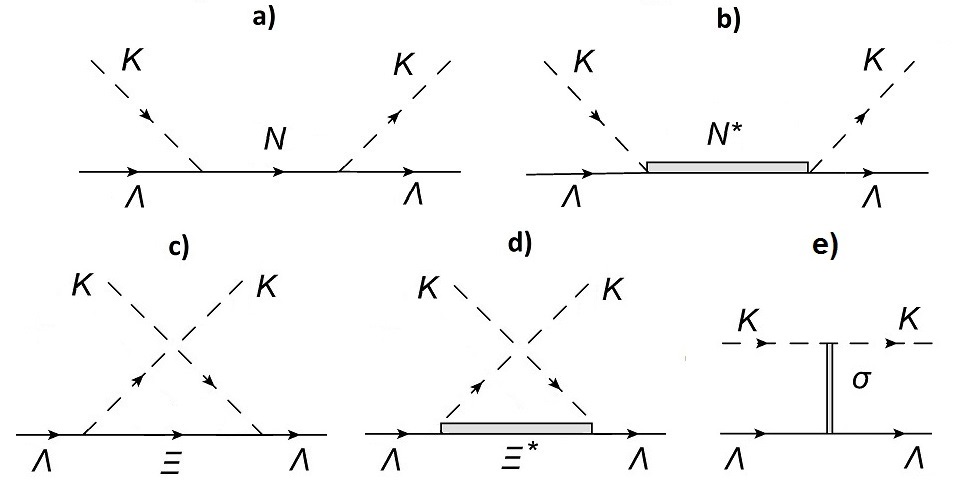}
 \caption{Diagrams for the $K\Lambda$ interation}\label{fig1}
\end{figure}
 
In Figure $\ref{fig1}$ the diagrams considered to describe the $K\Lambda$ interaction
are shown. The particles that will be considered in our calculations are listed in 
Tab.\ref{tb1} \cite{pdg}.
\begin{table}[!htb]
\begin{ruledtabular}
  \begin{tabular}{lccc} 
 & $J^\pi$ & $I$ &$Mass$ ($MeV$) \\ \hline
$N$ & $1/2^+$&1/2&938\\ 
$N(1650)$& $1/2^-$&1/2&1650\\ 
$N(1710)$ &$1/2^+$&1/2 &1710\\ 
$N^*(1875)$& $3/2^-$&1/2 &1875\\ 
 $N^*(1900)$&$3/2^+$&1/2&1900\\ 
$\Xi$ & $1/2^+$&1/2&1320\\ 
 $\Xi^*(1820)$& $3/2^-$&1/2&1820 \\
 \end{tabular}
 \caption{Particles considered in the $K\Lambda$ interaction}\label{tb1}
\end{ruledtabular}
\end{table}
 For the calculation of the contribution of particles with spin-1/2 ($N$ and $\Xi$) in the intermediate state
(Figure $\ref{fig1}a$ and $\ref{fig1}c$), the Lagrangian of interaction is \cite{BH}, \cite{sant}
\begin{equation}
\label{eq}
\mathcal{L}_{ \Lambda K B}= \frac{g_{\Lambda KB}}{2m_{\Lambda}}\big(\overline{B}\gamma_\mu\gamma_5\Lambda\big)\partial^\mu\phi\ ,
\end{equation}
where $\phi$ represents the kaon field, $B$ the baryon field, 
and $\Lambda$, the hyperon field, with mass $m_\Lambda$.

Calculating the Feynman diagrams and comparing with (\ref{eq20}) we find the amplitudes for the $N$ spin-1/2
exchange
\begin{eqnarray}
\label{eqc8}
A_N&=&\frac{g^2_{\Lambda KN}}{4m_\Lambda^2}(m_N+m_\Lambda)\bigg(\frac{s-m_\Lambda^2}{s-m_N^2}\bigg)\ ,\\
\label{eqc9}
B_N&=&-\frac{g^2_{\Lambda KN}}{4m_\Lambda^2}\bigg[\frac{2m_\Lambda(m_\Lambda+m_N)+s-m_\Lambda^2}{s-m_N^2}\bigg]   \   .
\end{eqnarray}  
For the contribution of the $\Xi$(1320) in the diagram of Fig.\ref{fig1}$c$ we
can calculate $A_\Xi$ and $B_\Xi$ by means of the following replacements
in eq. (\ref{eqc8}) and (\ref{eqc9}): 
$N\rightarrow \Xi$, $g^2_{\Lambda KN}\rightarrow -g^2_{\Lambda K\Xi}$ and $s\rightarrow u$, where $s$ and $u$ are the Mandelstam variables (defined in the appendix) and 
$g_{\Lambda KN(\Xi)}$ are the coupling constants obtained in \cite{sant}.

In a similar way, we adapted the Lagrangian for the interaction with
spin-3/2 resonances (Figure $\ref{fig1}$$b$ and 1$d$)
\begin{equation}
\label{eq}
\mathcal{L}_{ \Lambda KB^*}=  g_{ \Lambda KB^*}\Big\{\overline{B}^{*\mu}\big[g_{\mu\nu}-(Z+1/2)\gamma_\mu\gamma_\nu\big]\Lambda\Big\}\partial^\nu\phi\ .
\end{equation}

\noindent 
where, $B^*$ represents the spin-3/2 baryon field, and $Z$ is a free parameter.  

Calculating, the amplitudes representing
 the $N^*$ spin-3/2 (Fig. \ref{fig1}$b$) pole are
\begin{eqnarray}
\label{17}
A_{N^*}&=&\frac{g_{\Lambda KN^*}^2}{6}\bigg[\frac{2\hat{A}+3(m_\Lambda+m_{N^*})t}{m_{N^*}^2-s}+a_0\bigg]\ ,\\
\label{18}
B_{N^*}&=&\frac{g_{\Lambda KN^*}^2}{6}\bigg[\frac{2\hat{B}+3t}{m_{N^*}^2-s}-b_0\bigg]\ ,
\end{eqnarray}
where 
\begin{eqnarray}
\label{eq29'}
\hat{A}&=&3(m_\Lambda+m_{N^*})(\vec{q}_{N^*})^2\nonumber\\
&&+(m_{N^*}-m_\Lambda)(E_{N^*}+m_\Lambda)^2\ ,\\
\label{eq30'}
\hat{B}&=&3(\vec{q}_{N^*})^2-(E_{N^*}+m_\Lambda)^2\ ,\\
a_0&=&-\frac{(m_\Lambda+m_{N^*})}{m_{N^*}^2}\Big(2m_{N^*}^2+m_\Lambda m_{N^*}\nonumber\\
&&-m_\Lambda ^2+2m_K^2\Big)+\frac{4}{m_{N^*}^2}\Big[(m_{N^*}+m_\Lambda )Z\nonumber\\
&&+(2m_{N^*}+m_\Lambda )Z^2\Big]\Big[s-m_\Lambda^2\Big]\ ,
\end{eqnarray}
\begin{eqnarray}
\label{eq32'}
b_0&=&\frac{8}{m_{N^*}^2}\Big[(m_\Lambda ^2+m_\Lambda m_{N^*}-m_K^2)Z\nonumber\\
&&+(2m_\Lambda m_{N^*}+m_\Lambda ^2)Z\Big]+\frac{(m_\Lambda +m_{N^*})^2}{m_{N^*}^2}\nonumber\\
&&+\frac{4Z^2}{m_{N^*}^2}\Big[s-m_\Lambda^2\Big]\ ,
\end{eqnarray}
 where $t$, $\vec{q}_{N^*}$, $E_{N^*}$ are kinematical variables
defined in the appendix and $m_K$, $m_{N^*}$ are the kaon and the spin-3/2 resonance masses, respectively.

For the $\Xi^*$ spin-3/2 resonance contribution (Fig. \ref{fig1}$d$),
$A_{\Xi^*}$ and $B_{\Xi^*}$ are calculated making the substitutions
 $N^*\rightarrow\Xi^*$, $g^2_{\Lambda KN^*}\rightarrow g^2_{\Lambda K\Xi^*}$ and $s\rightarrow u$ in eqs. (\ref{17})-(\ref{eq32'}).

For the last diagram, Fig. \ref{fig1}$e$, that represents the scalar $\sigma$ meson exchange,
 a parametrization of the amplitude has been considered \cite{BH}-\cite{ccb}  
\begin{eqnarray}
\label{eq32}
A_\sigma&=&a+bt  \  ,\\
B_\sigma&=&0   \  ,
\label{eq331}
\end{eqnarray}
with $a=1.05 m_\pi^{-1}$, $b=-0.8m_\pi^{-3}$ and 
where $m_\pi$ is the pion mass \cite{pdg}.
Some discussions about this term may be found in \cite{cm}, \cite{leut1}-\cite{r1}.

\begin{figure}[!htb]
 \includegraphics[width=0.45\textwidth]{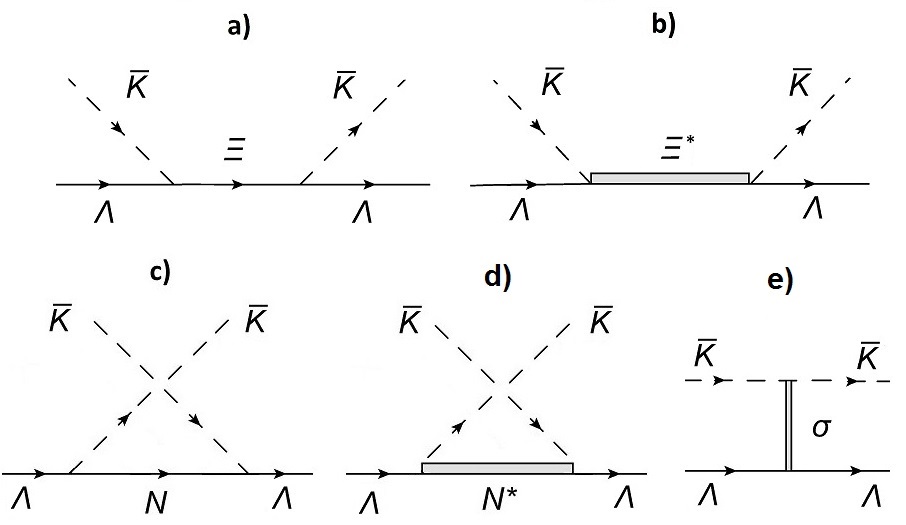}
 \caption{Diagrams for the $\overline K\Lambda$ interaction}\label{fig4}
\end{figure}

The $\overline K\Lambda$ interaction  may be studied
 exactly in same way that it has been done in the study of the
$K\Lambda$ interactions. 
Now we have the contributions presented in Figure $\ref{fig4}$, where the Lagrangians take
into account the $N$, $\Xi$, $\Lambda$ and $\phi'$ (representing the antikaon) fields.
\noindent

In this case the expressions (\ref{eqc8}), (\ref{eqc9}) and (\ref{17})-(\ref{eq32'})
may be used with the substitutions $N\rightarrow\Xi$, ${N^*}\rightarrow{\Xi^*}$,
 $m_{\overline{K}} = m_K$. For the  diagrams shown in Fig. \ref{fig4}$c$ and \ref{fig4}$d$
we make $\Xi\rightarrow N$, ${\Xi^*}\rightarrow{N^*}$ and $s\rightarrow u$
 and take into account the dominant resonances $N$(938), $N$(1650)  and $N^*$(1900). 

The parameters used in the formulation of
 the $K\Lambda$ and $\overline K\Lambda$ interaction are shown in Tab. \ref{tb2} \cite{sant}, \cite{pdg}.
\begin{table}[!htb]
\begin{ruledtabular}
  \begin{tabular}{cl}  
$m_\pi$& 140 $MeV$ \\
$m_K$& 496 $MeV$ \\ 
$m_\Lambda$& 1116 $MeV$ \\ 
$Z$& $-0.5$  \\ 
$g_{\Lambda KN}$ & 11.50\\
$g_{\Lambda KN(1650)}$ & 9.90 $GeV^{-1}$\\
$g_{\Lambda KN(1710)}$ & 5.20 $GeV^{-1}$\\
$g_{\Lambda KN^*(1875)}$& 0.53 $GeV^{-1}$ \\
$g_{\Lambda KN^*(1900)}$ & 2.60 $GeV^{-1}$\\
$g_{\Lambda K\Xi}$ & 0.24 \\
$g_{\Lambda K\Xi^*(1820)}$& 1.80 $GeV^{-1}$\\ 
 \end{tabular}
 \caption{Parameters for the $K\Lambda$ interaction}\label{tb2}
\end{ruledtabular}
\end{table}

The coupling constants shown in Tab. \ref{tb2}
have been determined in \cite{sant}. The
$g_{\Lambda KN}$ and $g_{\Lambda K\Xi}$ couplings
have been determined using $SU(3)$ \cite{swart}-\cite{stri} and the resonance couplings by comparing
our results with the Breit-Wigner expression. 

In the $K\Sigma$ scattering the interacting particles  have isospin 1/2 and 1, 
so, we have two possible total isospin states, 1/2 and 3/2, which allow the exchange of 
$\Delta$ particles too.
 
\begin{figure}[!htb]
 \includegraphics[width=0.50\textwidth]{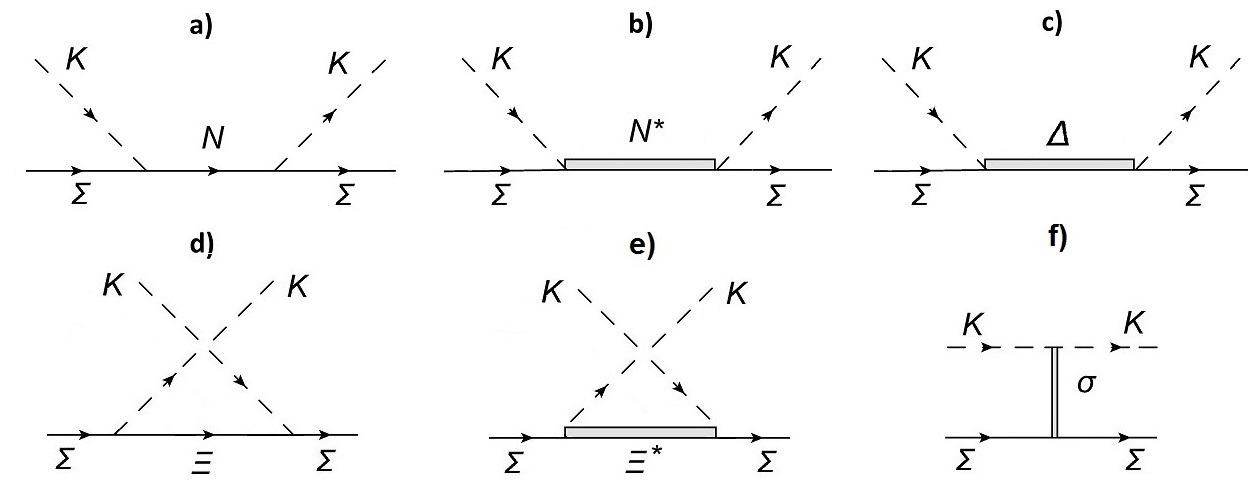}
 \caption{Diagrams for the $K\Sigma$ interaction}\label{ksig}
\end{figure}

The scattering amplitude has the general form
\begin{eqnarray}
T_{K\Sigma}^{\beta\alpha}&=&\bar{u}(\vec{p'})\bigg\{\Big[A^++\Big(\frac{\slashed k+\slashed k'}{2}\Big)B^+\Big]\delta^{\alpha\beta}\nonumber\\
&&+\Big[A^-+\Big(\frac{\slashed k+\slashed k'}{2}\Big)B^-\Big]i\epsilon^{\beta\alpha c}\tau_c\bigg\}u(\vec{p})\ ,
\label{eq}
\end{eqnarray} 
where we use the projection operators
\begin{eqnarray}
\label{eq37}
P^{\beta\alpha}_{\frac{1}{2}}=\frac{1}{3}\delta^{\beta\alpha}+\frac{i}{3}\epsilon^{\beta\alpha c}\tau_c \ ,\\
P^{\beta\alpha}_{\frac{3}{2}}=\frac{2}{3}\delta^{\beta\alpha}-\frac{i}{3}\epsilon^{\beta\alpha c}\tau_c\ ,
\label{eq381}
\end{eqnarray}
and the indices $\alpha$ and $\beta$ are relative to
the initial and final isospin states of the $\Sigma$. 
\begin{table}[!htb]
\begin{ruledtabular}
  \begin{tabular}{lccc}  
 & $J^\pi$ & $I$ &$Mass$ ($MeV$) \\ \hline
$N$ & $1/2^+$&1/2&938\\ 
$N(1710)$ &$1/2^+$&1/2 &1710\\ 
$N^*(1875)$& $3/2^-$&1/2 &1875\\ 
 $N^*(1900)$&$3/2^+$&1/2&1900\\ 
$\Delta(1920)$& $3/2^+$&3/2&1920\\ 
$\Xi$ & $1/2^+$&1/2&1320\\ 
 $\Xi^*(1820)$& $3/2^-$&1/2&1820 \\
 \end{tabular}
 \caption{Resonances of the $K\Sigma$ interaction }\label{tb3} 
\end{ruledtabular}
\end{table}


The contributing diagrams are shown
in Fig. $\ref{ksig}$ and the particles to be considered
 in Tab. \ref{tb3}. The Lagrangian for the exchange of spin-1/2 particles now become,
\begin{equation}
\label{eq39}
\mathcal{L}_{\Sigma K B}= \frac{g_{\Sigma KB}}{2m_{\Sigma}}\big(\overline{B}\gamma_\mu\gamma_5\vec{\tau}.\vec{\Sigma}\big)\partial^\mu\phi  \  ,
\end{equation}
and for spin-3/2 particles 
\begin{equation}
\label{eq40}
\mathcal{L}_{\Sigma KB^*}=  g_{ \Sigma KB^*}\Big\{\overline{B}^{*\mu}\big[g_{\mu\nu}-(Z+1/2)\gamma_\mu\gamma_\nu\big]\vec{Q}.\vec{\Sigma}\Big\}\partial^\nu\phi  \  ,
\end{equation}
where $\vec{Q}$ is the $\vec{M}$ matrix, that combines a isospin 1/2 baryon and a $\Delta$ ($I=3/2$)
into a isospin 1 state, or the  $\vec{\tau}$ matrix, that combines
two isospin 1/2 particles, $N^*$ and $\Xi^*$, into a isospin 1 state.

The resulting amplitudes for spin-1/2 $N$ particles in the intermediate state are
 (Fig. \ref{ksig}$a$)
\begin{eqnarray}
\label{eq31}
A_N^+&=&\frac{g^2_{\Sigma KN}}{4m_\Sigma^2}(m_N+m_\Sigma)\bigg(\frac{s-m_\Sigma^2}{s-m_N^2}\bigg)\ ,\\
B_N^+&=&-\frac{g^2_{\Sigma KN}}{4m_\Sigma^2}\bigg[\frac{2m_\Sigma(m_\Sigma+m_N)+s-m_\Sigma^2}{s-m_N^2}\bigg]\ ,\\
A_N^-&=&\frac{g^2_{\Sigma KN}}{4m_\Sigma^2}(m_N+m_\Sigma)\bigg(\frac{s-m_\Sigma^2}{s-m_N^2}\bigg)\ ,\\
\label{eq34}
B_N^-&=&-\frac{g^2_{\Sigma KN}}{4m_\Sigma^2}\bigg[\frac{2m_\Sigma(m_\Sigma+m_N)+s-m_\Sigma^2}{s-m_N^2}\bigg]\ ,
\end{eqnarray}
and to determine the amplitudes of
the diagram \ref{ksig}$d$, that represents a spin-1/2 $\Xi$ in the intermediate state, 
we proceed in the same way that it has been done in the study of the
$K\Lambda$ interaction.

For the spin-3/2 $N^*$ particles (Fig. \ref{ksig}$b$) we have
\begin{eqnarray}
\label{eq35}
A_{N^*}^+&=&\frac{g_{\Sigma KN^*}^2}{6}\bigg[\frac{2\hat{A}+3(m_\Sigma+m_{N^*})t}{m_{N^*}^2-s}+a_0\bigg]\ ,\\
B_{N^*}^+&=&\frac{g_{\Sigma KN^*}^2}{6}\bigg[\frac{2\hat{B}+3t}{m_{N^*}^2-s}-b_0\bigg]\ ,\\
A_{N^*}^-&=&\frac{g_{\Sigma KN^*}^2}{6}\bigg[\frac{2\hat{A}+3(m_\Sigma+m_{N^*})t}{m_{N^*}^2-s}+a_0\bigg]\ ,\\
\label{eq38}
B_{N^*}^-&=&\frac{g_{\Sigma KN^*}^2}{6}\bigg[\frac{2\hat{B}+3t}{m_{N^*}^2-s}-b_0\bigg]\ ,
\end{eqnarray}
where we use the precedent results ($\ref{eq29'}$)-($\ref{eq32'}$)
 replacing the $\Lambda$ hyperon for the $\Sigma$ hyperon. 
 In order to study 
 the spin-isospin-3/2 $\Delta$ resonance exchange in Fig. \ref{ksig}$c$, we make $N^*\rightarrow\Delta$ and $g^2_{\Sigma KN^*}/6\rightarrow g^2_{\Sigma K\Delta}/9$ to the $(+)$ amplitudes and $g^2_{\Sigma KN^*}/6\rightarrow g^2_{\Sigma K\Delta}/18$ to the $(-)$ amplitudes.

Finally for the interaction shown in Fig. \ref{ksig}$e$, that considers a
spin-3/2 $\Xi^*$, the results of the $K\Lambda$ interaction may be used with
the correct adaptations and for Fig. \ref{ksig}$f$ we use the (\ref{eq32}) and (\ref{eq331}).

Thus, to calculate the observables for each reaction
 we use ($\ref{eq37}$) and ($\ref{eq381}$), resulting in the amplitudes 

\begin{eqnarray}
\label{eq55}
A^{\frac{1}{2}}&=&A^++2A^-\ ,\\
\label{eq56}
B^{\frac{1}{2}}&=&B^++2B^-\ ,\\
\label{eq57}
A^{\frac{3}{2}}&=&A^+-A^-\ ,\\
\label{eq58}
B^{\frac{3}{2}}&=&B^+-B^-\ .
\end{eqnarray} 
\noindent

Using the isospin formalism for the elastic and charge exchange scattering, 
we can determine the amplitudes for the reactions (that we name $C_i$, for simplicity)
\begin{eqnarray}
\label{eqc1}
\left\langle\Sigma^+K^+|T|\Sigma^+K^+ \right\rangle&=&\nonumber \left\langle\Sigma^-K^0|T|\Sigma^-K^0 \right\rangle\\ 
&=&T_{\frac{3}{2}}\equiv C_1\ ,\\\nonumber
\label{eq:}
\left\langle\Sigma^+K^0|T|\Sigma^+K^0 \right\rangle&=&\left\langle\Sigma^-K^+|T|\Sigma^-K^+ \right\rangle\\
&=&\frac{1}{3}T_{\frac{3}{2}}+\frac{2}{3}T_{\frac{1}{2}}\equiv  C_2\ ,\\ \nonumber
\label{eq:}
\left\langle\Sigma^0K^0|T|\Sigma^0K^0 \right\rangle&=&\left\langle\Sigma^0K^+|T|\Sigma^0K^+ \right\rangle\\
&=&\frac{2}{3}T_{\frac{3}{2}}+\frac{1}{3}T_{\frac{1}{2}}\equiv  C_3\ ,\\ \nonumber
\label{eq:}
\left\langle\Sigma^0K^0|T|\Sigma^-K^+ \right\rangle&=&\left\langle\Sigma^+K^0|T|\Sigma^0K^+ \right\rangle\\ \nonumber
&=&\left\langle\Sigma^-K^+|T|\Sigma^0K^0 \right\rangle\\\nonumber
&=&\left\langle\Sigma^0K^+|T|\Sigma^+K^0 \right\rangle\\
&=&\frac{\sqrt{2}}{3}\Big(T_{\frac{3}{2}}-T_{\frac{1}{2}}\Big)\equiv  C_4\ ,
\label{eq:}
\end{eqnarray}
and
with these amplitudes we can calculate the angular distribution and the polarization to each isospin channel. 

\begin{figure}[!htb]
 \includegraphics[width=0.45\textwidth]{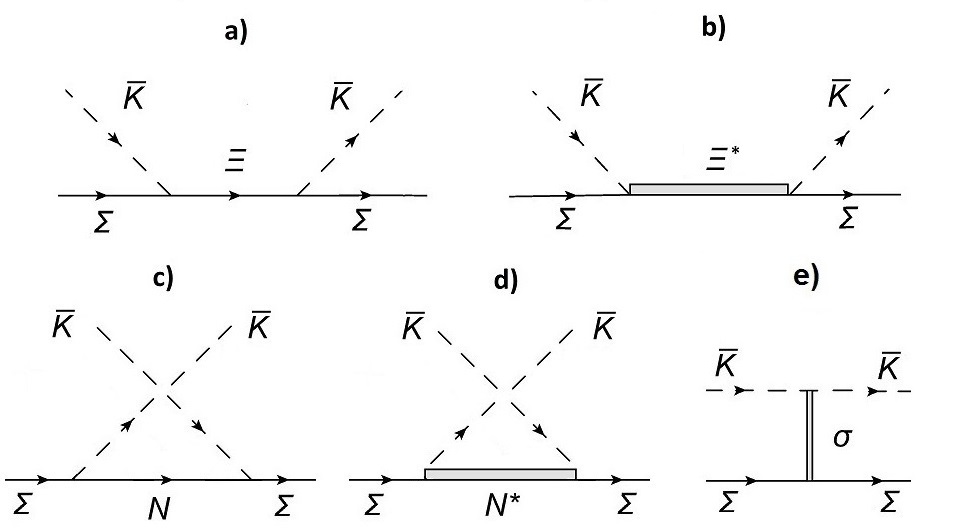}
 \caption{Diagrams for the $\overline K\Sigma$ interaction}\label{kbarsig}
\end{figure}

The last case to be studied is the $\overline{K}\Sigma$ interaction and we will
we proceed in the same way as we have done for the $K\Sigma$ interaction.  The diagrams to be considered are shown in Fig.\ref{kbarsig}.

 The Lagrangians $(\ref{eq39})$ and $(\ref{eq40})$
 may be used with
 $\phi\rightarrow\phi'$, $N\rightarrow\Xi$ and $N^*\rightarrow\Xi^*$, resulting in the same structure of the amplitudes given in eqs. $(\ref{eq31})-(\ref{eq38})$. 

The reactions to be studied (elastic and charge exchange) are
\begin{eqnarray}
\label{eqd1}
\left\langle \overline K^0\Sigma^+|T| \overline K^0\Sigma^+ \right\rangle\nonumber&=&\left\langle K^-\Sigma^-|T|  K^-\Sigma^- \right\rangle\\
&=&T_{\frac{3}{2}}\equiv D_1 \ ,\\\nonumber
\left\langle\Sigma^+K^-|T|\Sigma^+K^- \right\rangle&=&\left\langle\Sigma^-\overline{K}^0|T|\Sigma^-\overline{K}^0 \right\rangle\\
&=&\frac{1}{3}T_{\frac{3}{2}}+\frac{2}{3}T_{\frac{1}{2}}\equiv D_2\ ,\\\nonumber
\label{eq:}
\left\langle\Sigma^0\overline{K}^0|T|\Sigma^0\overline{K}^0 \right\rangle&=&\left\langle\Sigma^0K^-|T|\Sigma^0K^- \right\rangle\\
&=&\frac{2}{3}T_{\frac{3}{2}}+\frac{1}{3}T_{\frac{1}{2}}\equiv D_3\ ,\\\nonumber
\label{eq:}
\left\langle\Sigma^0K^-|T|\Sigma^-\overline K^0 \right\rangle&=&\nonumber\left\langle\Sigma^+K^-|T|\Sigma^0\overline K^0 \right\rangle\\
&=&\nonumber\left\langle\Sigma^-\overline K^0|T|\Sigma^0K^- \right\rangle\\
&=&\nonumber\left\langle\Sigma^0\overline K^0|T|\Sigma^+K^- \right\rangle\\
&=&\frac{\sqrt{2}}{3}\Big(T_{\frac{3}{2}}-T_{\frac{1}{2}}\Big)\equiv  D_4\ .
\label{eq:}
\end{eqnarray}

In the diagrams of Fig. \ref{kbarsig}$c$ and \ref{kbarsig}$d$ we have considered 
 $N(938)$, $N(1710)$ and $N^*(1900)$. 

 The parameters for the $K\Sigma$ and $\overline K\Sigma$ interactions
 are given in Tabs. \ref{tb2} and \ref{tb4} \cite{sant}. 

\begin{table}[!htb]
\begin{ruledtabular}
  \begin{tabular}{cl}  
$m_\Sigma$& 1190 $MeV$ \\ 
$g_{\Sigma KN}$ & 6.90\\ 
$g_{\Sigma KN(1710)}$ & 6.85 $GeV^{-1}$\\
$g_{\Sigma KN^*(1875)}$& 0.70 $GeV^{-1}$ \\
$g_{\Sigma KN^*(1900)}$ & 1.30 $GeV^{-1}$\\
$g_{\Sigma K\Delta(1920)}$& 1.70 $GeV^{-1}$\\
$g_{\Sigma K\Xi}$ & 13.40 \\ 
$g_{\Sigma K\Xi^*(1820)}$& 1.80 $GeV^{-1}$\\ 

 \end{tabular}
 \caption{Parameters for the $K\Sigma$ interaction}\label{tb4}
\end{ruledtabular}
\end{table}
\section{Results}
In this section we present all the results for the $K\Lambda$, $\overline K\Lambda$, $K\Sigma$ and 
$\overline K\Sigma$ scattering. 
In the polarization graphics we use a gap of 10 $MeV$ between the lines and 
in the angular distribution ones we plot continuous surfaces. 
These observables have been calculated as functions of $k$, the incident momentum 
absolute value and $x=\cos\theta$, where $\theta$ is the scattering angle,
in the center-of-mass frame as defined in the previous sections. 

\begin{figure}[!htb]
\includegraphics[width=0.410\textwidth]{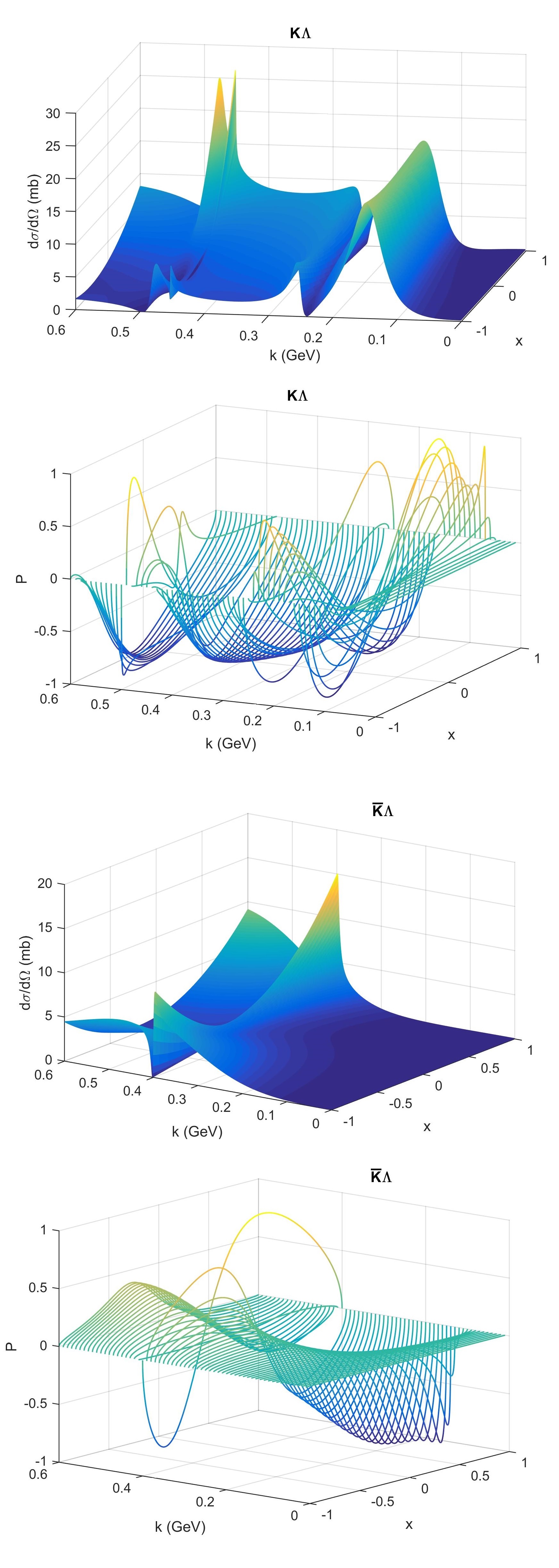}
\caption{Angular Distribution and Polarization in the $K\Lambda$ and $\overline K\Lambda$ scattering}\label{fig5}
\end{figure}

In Fig.\ref{fig5} the resulting angular distribution and the polarization in the $K\Lambda$ and $\overline K\Lambda$ scattering 
are presented.

\begin{figure}[!htb]
\includegraphics[width=0.5\textwidth]{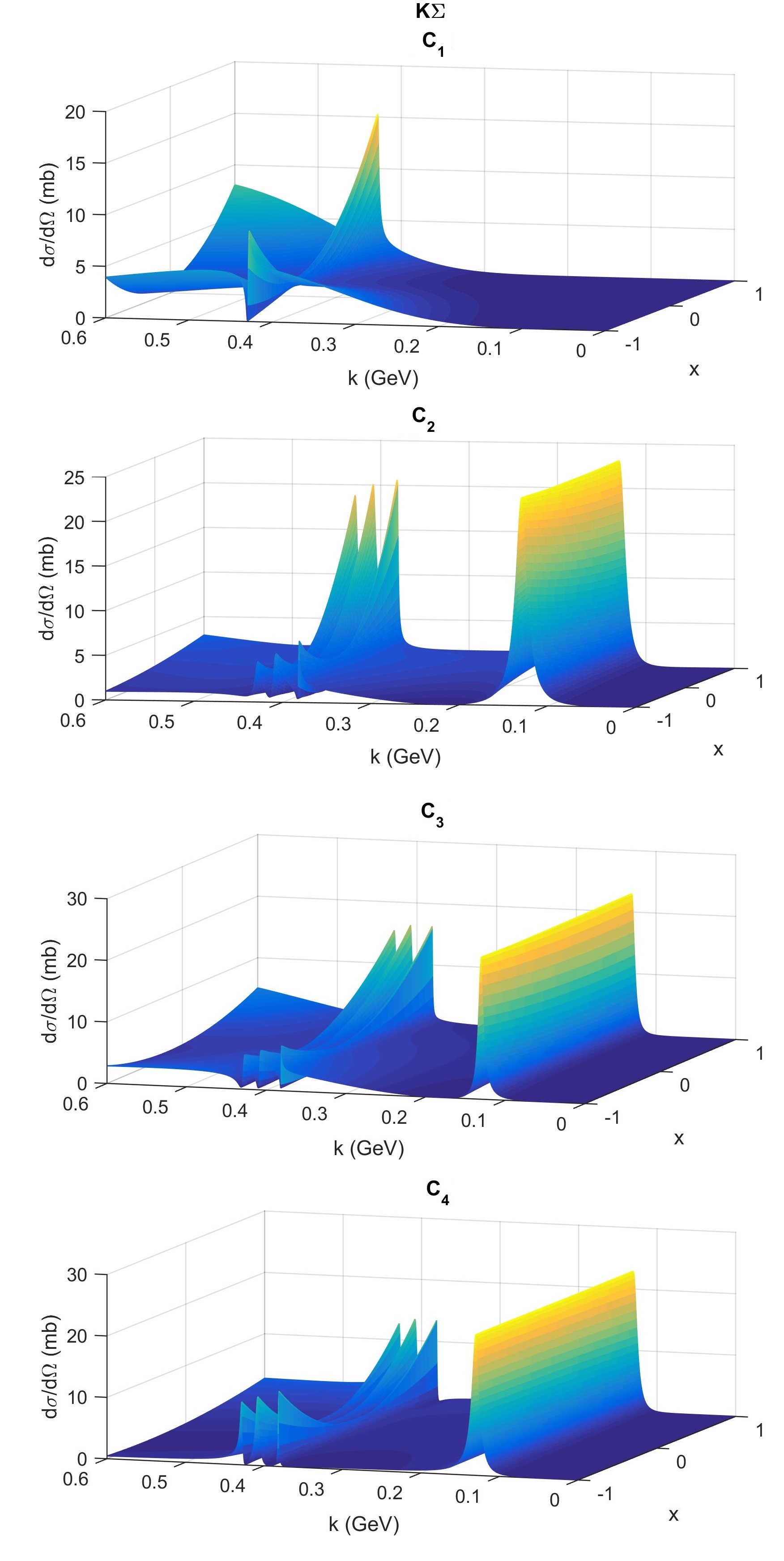}
\caption{Angular Distribution in the $K\Sigma$ scattering}\label{fig7}
\end{figure}

\begin{figure}[!htb]
\includegraphics[width=0.5\textwidth]{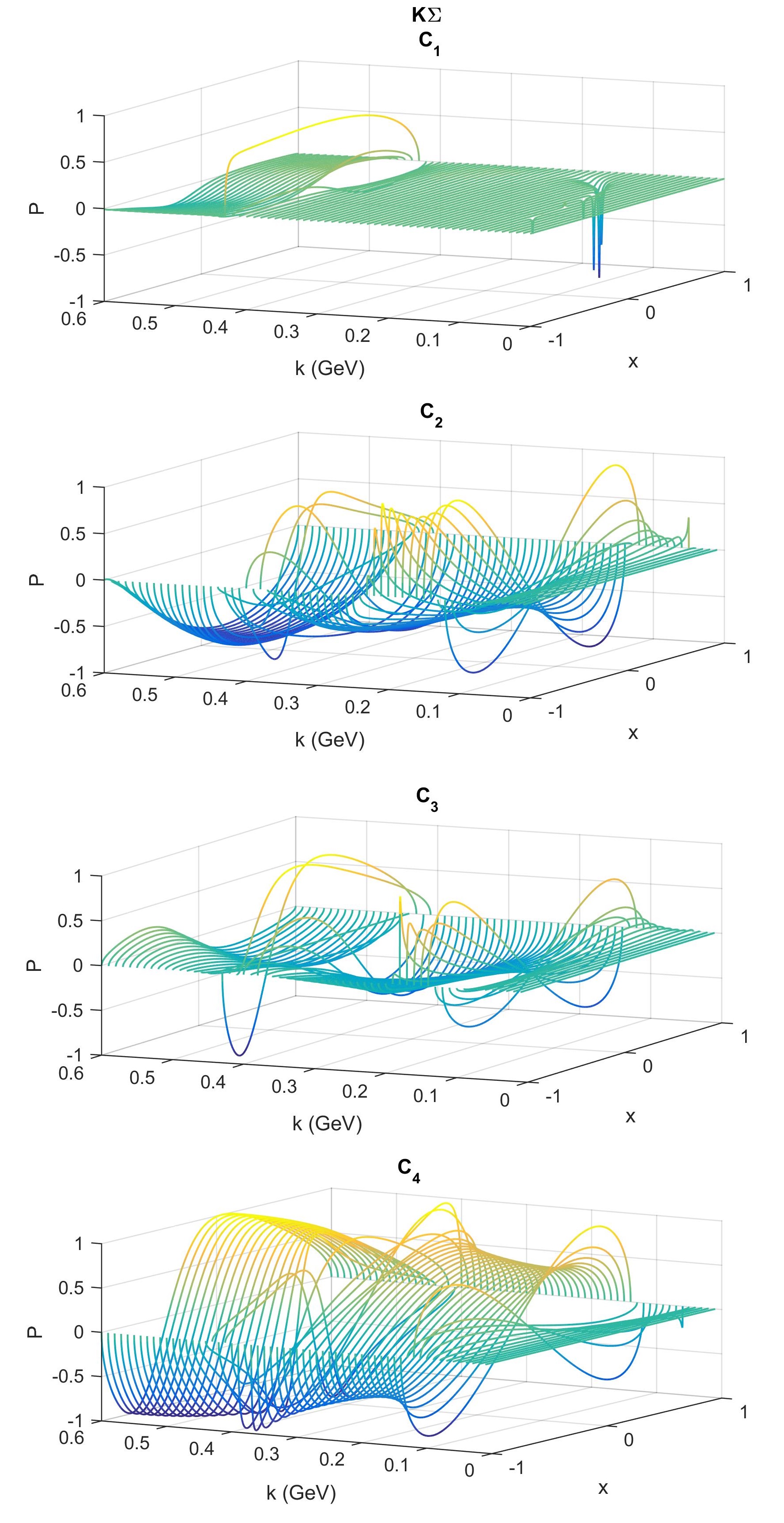}
\caption{Polarization in the $K\Sigma$ scattering}\label{fig8}
\end{figure}

\begin{figure}[!htb]
\includegraphics[width=0.5\textwidth]{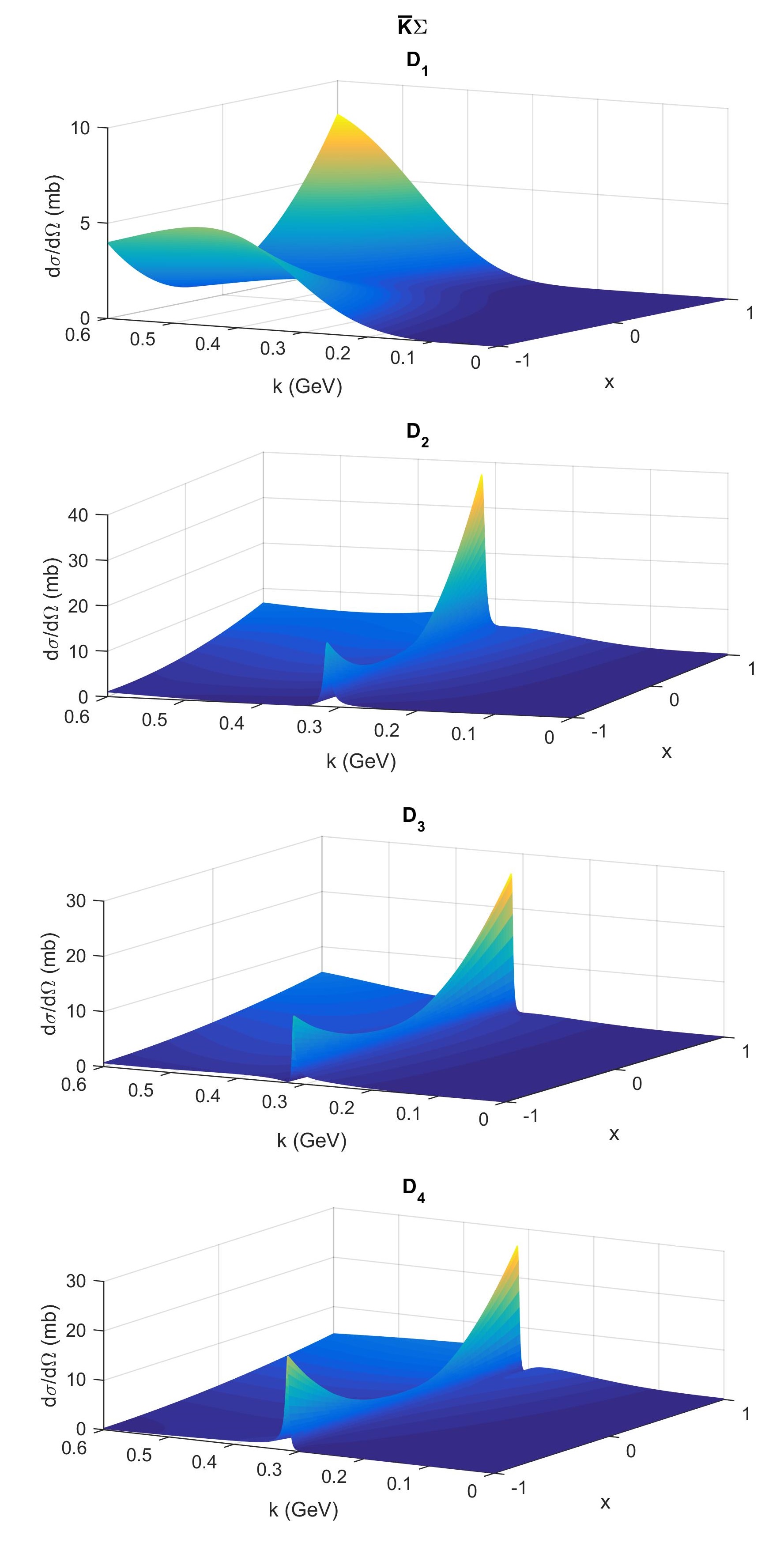}
\caption{Angular Distribution in the $\overline K\Sigma$ scattering}\label{fig9}
\end{figure}

\begin{figure}[!htb]
\includegraphics[width=0.5\textwidth]{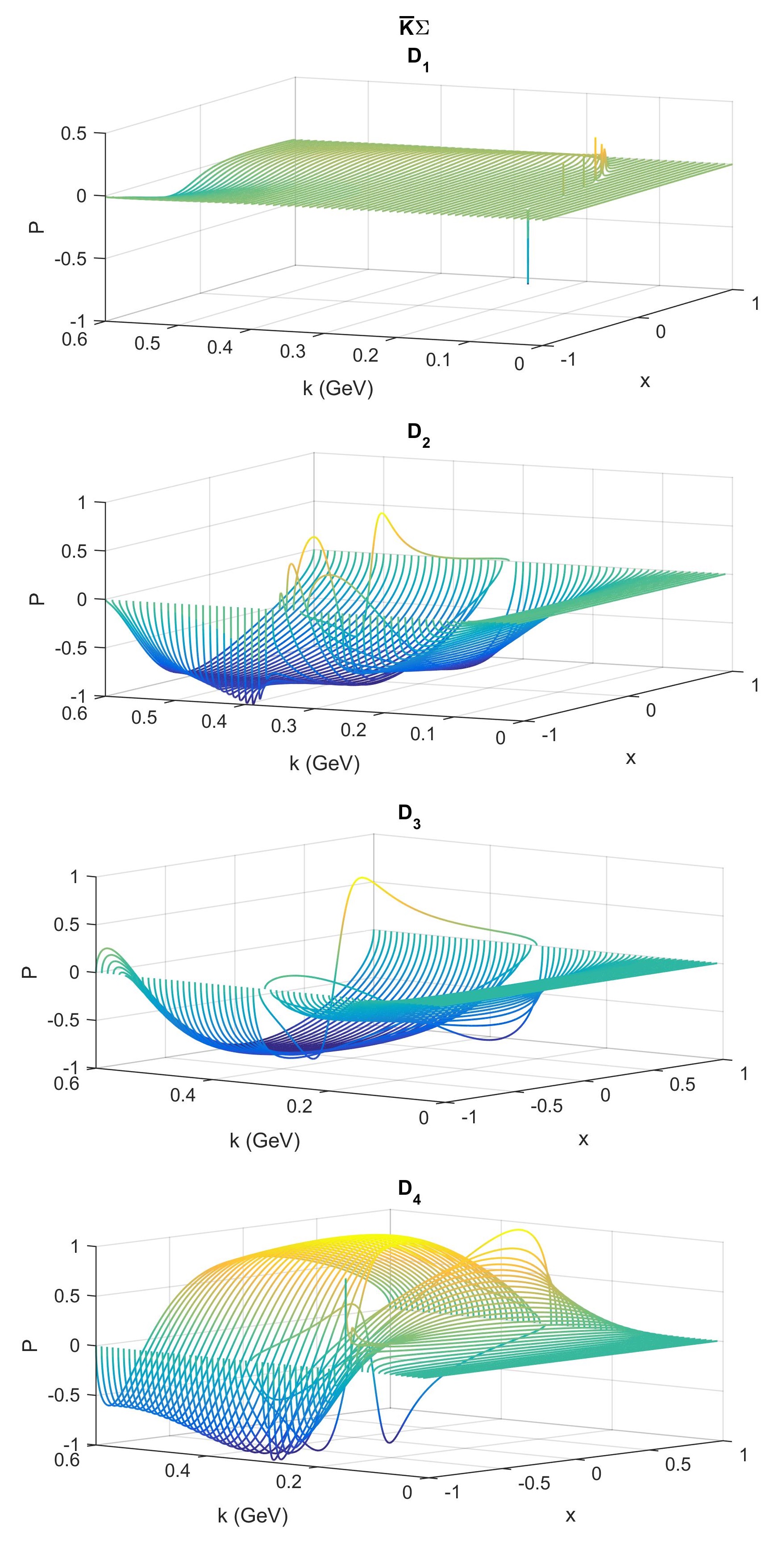}
\caption{Polarization in the $\overline K\Sigma$ scattering}\label{fig10}
\end{figure}

The angular distributions in the $K\Sigma$ interactions
for all possible channels are shown in Fig.\ref{fig7} and the polarizations in Fig.\ref{fig8}. 
Finally, the angular distributions and polarizations for the $\overline K\Sigma$ interactions are shown in Fig.\ref{fig9} and \ref{fig10} respectively. 

As we can see in the figures, a basic feature is the dominance of the resonances in the cross sections
at low energies, fact that is similar to the observed behavior in the very well known
pion-nucleon interactions, where the $\Delta$ dominates the cross sections, specially in the 
spin-3/2 and isospin-3/2 channel. When observing the polarizations, we see that the
reactions labeled by $C_1$ and $D_1$, relative to 
isospin 3/2 channels that are $K^+\Sigma^+\rightarrow K^+\Sigma^+$, 
$K^-\Sigma^-\rightarrow K^-\Sigma^-$ 
(and others, see eq. (\ref{eqc1}) and (\ref{eqd1})), 
it may be very small. In the other channels, 
large values of the polarization may be observed, it may be positive or negative,
and in general, oscillates.


\section{Discussion}

In this paper we have calculated the angular distributions and polarizations for many
reactions considering kaon-hyperon and antikaon hyperon interactions. We have considered a model
based in effective nonlinear chiral lagrangians, previously used to describe
the pion-hyperon \cite{BH} and pion-nucleon \cite{manc} interactions.

As we pointed before, an important aspect is that at low energies the resonances play an important
role in the cross sections. The polarizations in general are oscillating functions, that may be small,
as it can be seen in the reactions 
labeled by $C_1$ and $D_1$ (Fig. \ref{fig8} and 
Fig. \ref{fig10}), but also
may be consistenly large, as for example in the reactions
represented by $D_4$ shown in Fig. \ref{fig10}.

Despite the fact that this kind of reactions are very important in order to
understand the basic proprieties of the hyperon physics, we imagine that, maybe
due to technical difficulties, in a near fucture this kind of observable
will not be measured experimentally. A possible way to determine 
some phase-shifts is a procedure similar to the one used in the 
HyperCP experiment \cite{hyper1}, \cite{hyper2}, where $\delta_S-\delta_P$ has been determined 
for the $\pi\Lambda$ interaction
observing the production and decay of the $\Xi$ hyperon in high energy experiments.
The results presented in \cite{BH} are in good accord with these data, and then
validate the model proposed in the study of the meson-hyperon interactions.
We expect that this kind of question motivates future experiments.

Although, as it has been said before, the main motivation for this work is
the polarization of hyperon produced in high energy collisions. As it has been
shown in \cite{cy}-\cite{ccb2}, the low energy meson-hyperon interactions are key 
elements, as far as the produced hyperons interact with the surrounding medium
before the detection, and these interactions may affect the final hyperon
polarization. In \cite{cy}-\cite{ccb2}, only the pion-hyperon interactions have been 
considered and this fact motivated this work. So, a natural improvement of the
model is the inclusion of the kaon-hyperon interactions.
Observing the polarization results
in Fig.\ref{fig5}-\ref{fig10} we may conclude that they can have some effect
in high energy processes if the final-state interactions are considered, 
and these corrections may be interpreted as experimental
verifications of the results presented in this work. These calculations will be shown
in future works.


\section{Acknowledgments}

This study has been partially supported by the Coordena\c c\~ao
de Aperfei\c coamento de Pessoal de N\'{\i}vel Superior
(CAPES) – Finance Code 001.

\section{Appendix}
Considering a process where
 $p$ and $p'$ are
the initial and final hyperon four-momenta, $k$ and $k'$ are the initial and final meson 
four-momenta, the Mandelstam variables are given by
\begin{eqnarray*}
s&=&(p+k)^2=(p'+k')^2=m^2+m_K^2+2Ek_0-2\vec{k}.\vec{p}\ ,\\
u&=&(p'-k)^2=(p-k')^2=m^2+m_K^2-2Ek_0-2\vec{k}'.\vec{p}\ ,\\
t&=&(p-p')^2=(k-k')^2=2(\vec{k})^2x-2(\vec{k})^2  \   .
\end{eqnarray*}
In the center-of-mass frame, the energies will be
defined as
\begin{eqnarray*}
&&k_0=k'_0=\sqrt{|\vec{k}|^2+m_K^2}\\
&&E=E'=\sqrt{|\vec{k}|^2+m^2}
\label{eq:}
\end{eqnarray*}
where $m$ and $m_K$ are the hyperon mass and the kaon mass, respectively.

We also define the variable
\[
x=\cos\theta
\]
where $\theta$ is the scattering angle.
For the energy and for the 3-momentum of the intermediary particles we have the relations
\begin{eqnarray*}
&&(E_{B^*}\pm m_\Lambda)=\frac{(m_{B^*}\pm m_\Lambda)^2-m_K^2}{2m_{B^*}} \ ,\\
&&(\vec{q}_{B^*})^2=E_{B^*}^2- m_\Lambda^2=(E_{B^*}+ m_\Lambda)(E_{B^*}- m_\Lambda)\ .
\end{eqnarray*}
where $E_{B^*}$ and $\vec{q}_{B^*}$  are the energy and the momentum of the  intermediary baryon $B^*$ in the center-of-mass frame, respectively.



\end{document}